\documentclass[12pt,a4paper,final]{iopart}

\usepackage{iopams}  
\usepackage{graphicx, import}
\usepackage[breaklinks=true,colorlinks=true,linkcolor=blue,urlcolor=blue,citecolor=blue]{hyperref}

\newcommand{\be}{\begin{equation}}
\newcommand{\ee}{\end{equation}}
\newcommand{\bea}{\begin{eqnarray}}
\newcommand{\eea}{\end{eqnarray}}

\begin{document}

\title[DQD-QPC: A stoch. thermo. approach]{Double quantum dot coupled to a quantum point contact: A stochastic thermodynamics approach}

\author{Gregory {Bulnes Cuetara} and Massimiliano Esposito}
\address{Complex Systems and Statistical Mechanics, Physics and Materials Science Research Unit, University of Luxembourg, L-1511 Luxembourg, Luxembourg}



\begin{abstract}
We study the nonequilibrium properties of an electronic circuit composed of a double quantum dot (DQD) channel coupled to a quantum point contact (QPC) within the framework of stochastic thermodynamics. We show that the transition rates describing the dynamics satisfy a nontrivial local detailed balance (LDB) and that the statistics of energy and particle currents across both channels obeys a fluctuation theorem (FT). We analyze two regimes where the device operates as a thermodynamic machine and study its output power and efficiency fluctuations. We show that the electrons tunneling through the QPC without interacting with the DQD have a strong effect on the device efficiency.
\end{abstract}



\section{Introduction}


Semiconducting multichannel circuits made of quantum dots and quantum point contacts are nowadays commonly devised and studied experimentally \cite{Schoelkopf_1998_Science, Lu_2003_Nature, Fujisawa_2004_Appliedphysicsletters, Bylander2005Nature, Gustavsson_2006_PhysicalReviewLetters, Gustavsson_2007_PhysicalReviewLetters, Pekola_2007_PhysicalReviewLetters, Kafanov_2009_PhysicalReviewLetters, Saira_2012_PhysicalReviewLetters, Koski_2013_NaturePhysics}. 
The progress in the control of electronic temperatures at the meso-scale \cite{Pekola_2007_PhysicalReviewLetters, Kafanov_2009_PhysicalReviewLetters} has for instance driven the experimental \cite{Tian_2007_Nature, Flipse_2014_Physicalreviewletters} and theoretical \cite{Humphrey_2002_Physicalreviewlettersa, Humphrey_2005_Physicalreviewlettersa, Esposito_2009_EPL, Sanchez_2011_PhysRevB, Esposito_2012_PhysicalReviewEa, Jordan_2013_PhysRevB, Brandner_2013_NewJournalofPhysics, Brandner_2013_Physicalreviewletters, Entin-Wohlman_2014_PhysicalReviewE, Whitney_2014_Physicalreviewletters} study of their thermoelectric properties. 
In the isothermal case, these circuits have also been used to probe the fluctuating properties of heat and matter transfers using counting statistics experiments \cite{Gustavsson_2006_PhysicalReviewLetters, Kung_2012_PhysicalReviewX, Saira_2012_PhysicalReviewLetters, Koski_2013_NaturePhysics}. 
A circuit of particular interest in that regard is the DQD channel probed by a QPC detector. It has been used to perform the bidirectional counting of single electrons in the DQD channel \cite{Fujisawa_2006_Science, Kung_2012_PhysicalReviewX} and to provide the first experimental verification of the current FT in mesoscopic physics \cite{Utsumi_2010_PhysicalReviewB}. Theoretically, several studies have analyzed the backaction of the QPC detector on the mean current of the DQD channel, as well as on its statistics \cite{Golubev_2011_PhysicalReviewB, Cuetara_2013_PhysicalReviewB, Ouyang_2010_PhysicalReviewB, Li_2013_Scientificreports}. The QPC detector was also shown to modify the thermodynamic affinity of the DQD channel while preserving the FT symmetry in the DQD circuit \cite{Cuetara_2013_PhysicalReviewB}. In this latter work, the tunneling events in the QPC were treated non perturbatively to account for possible high transparency and as a result, the combined DQD-QPC statistics was not accessible within this approach.

In this paper, we study the nonequilibrium thermodynamics of the DQD-QPC circuit using stochastic thermodynamics \cite{Seifert_2012_ReportsonProgressinPhysics, VandenBroeck_2014_ArXive-prints, Esposito_2012_PhysicalReviewE}. We consider the general case where the QPC and DQD reservoirs may be at a different temperatures and chemical potentials. The counting statistics of the energy and matter currents across both channels is calculated using the modified quantum master equation formalism \cite{Esposito_2009_ReviewsofModernPhysics} assuming weak coupling between the DQD and its reservoirs as well as between the reservoirs composing the QPC. Even though in the isothermal case this last assumption may seem more restrictive compared to the nonperturbative approach of Ref.\cite{Cuetara_2013_PhysicalReviewB, Schaller_OpenQuantumSystemsFarfromEquilibrium}, it enables us to analytically calculate the join distribution of the energy and matter currents across both channels, and to identify the entropy flows associated to the exchange processes at hand. As a result, we are able to derive a bivariate FT for the statistics of the energy and matter currents in both channels.

An interesting feature of this setup is that the microscopic processes associated to transitions in the DQD involve more than one reservoir at a time. In particular, transitions in the DQD induced by the QPC have corresponding transition rates proportional to the product of Fermi functions in both reservoirs of the QPC. The microscopic processes underlying such transitions involve the tunneling of an electron between the QPC reservoirs which exchanges a fraction of its energy with the DQD. As a result, the transition rates cannot be written anymore as a sum where each term only involves one single reservoir. Despite this non additivity of the rates, the local detailed balance (LDB) \cite{Esposito_2009_ReviewsofModernPhysics, Kubo_Statistical-mechanicaltheory} is shown to hold, and is explicitly written in terms of the fluxes of entropy from the reservoirs involved in the transitions.

Furthermore, since some electrons tunnel between the QPC reservoirs without exchanging energy with the DQD, they do not induce transitions in the DQD but need to be taken into account in the counting statistics. We show that their statistics is well described by the modified quantum master equation and corresponds to the Levitov-Lesovik formula \cite{Levitov_1993_JETPLETTERSC, Levitov_1996_JournalofMathematicalPhysics, Schoenhammer_2007_PhysicalReviewB, Klich_2002_eprintarXiv, Gogolin_2006_PhysicalReviewB} to second order perturbation theory in the coupling between the QPC reservoirs. Though such processes do not reveal themselves in the master equation for the DQD populations, they are shown to strongly impact the circuit performance.

We proceed by analyzing two different regimes where the circuit operates as a thermoelectric and as a current converter, respectively. We identify the optimal working conditions to reach large average output power and high macroscopic efficiency and study the statistical properties of the output power as well as of the efficiency, as recently proposed in Refs. \cite{Verley_2014_NatureCommunications, Verley_2014_ArXive-prints, Esposito_2015_PhysRevB}. 

The paper is organized as follows: The model is introduced in section \ref{hamiltonian}. In section \ref{countingstatistics}, we derive the modified quantum master equation to calculate the counting statistics of energy and matter currents, and analyze the microscopic processes contributing to the transition rates. In section \ref{noneqthermodynamics}, using the LDB property of the rates, we identify the entropy flows associated to each microscopic processes in the circuit, and the steady state FT is derived. The expressions for the average energy and matter currents as well as the average irreversible entropy production are also provided. The thermodynamic analysis of the circuit operating as a thermoelectric and as a current converter is done in section \ref{engines}. Conclusions are drawn in section \ref{summary}.


\section{Hamiltonian}\label{hamiltonian}

\begin{figure}[htbp]
\centerline{\includegraphics[width=10cm]{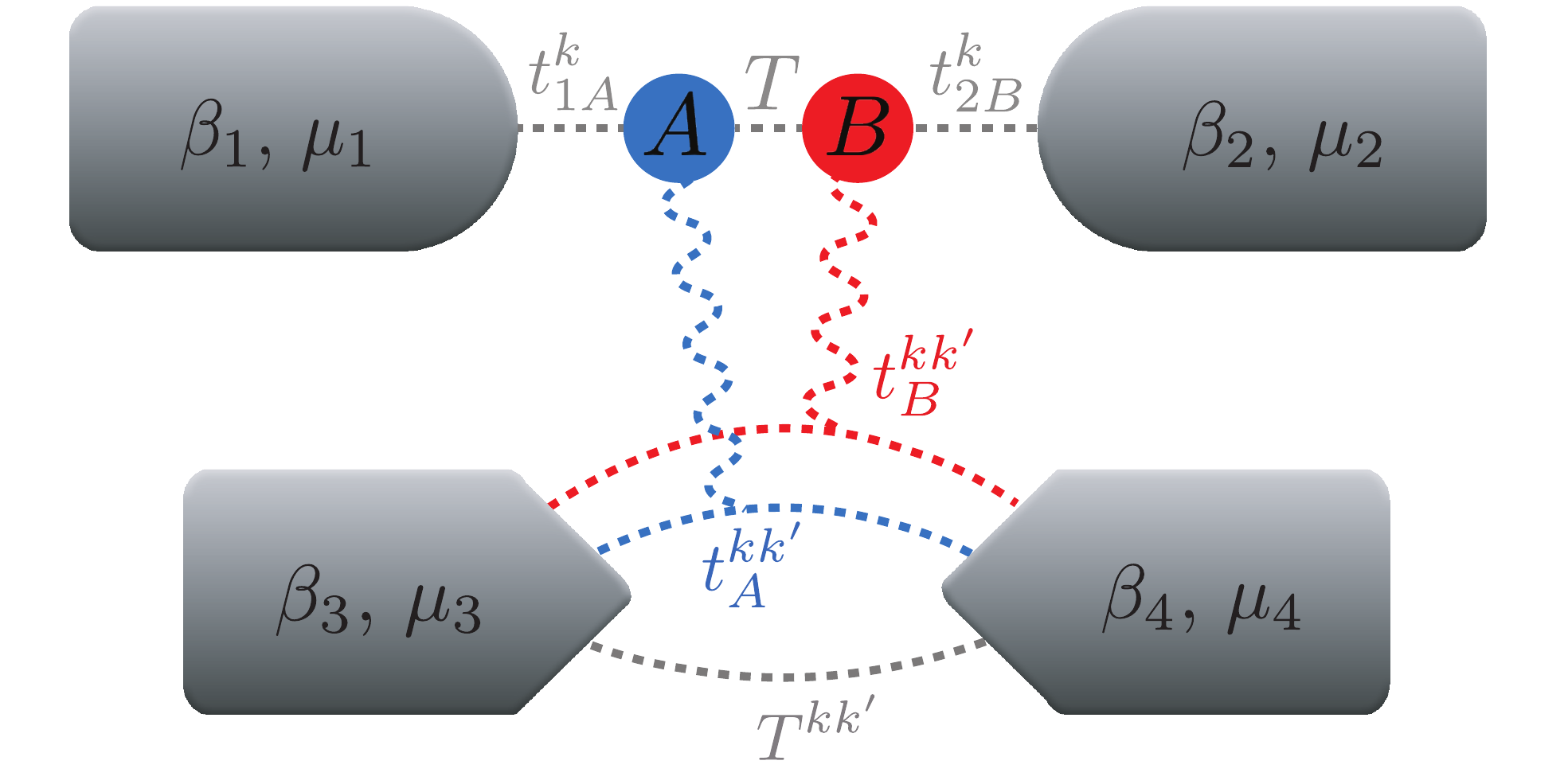}}
\caption{Double channel circuit made of a DQD and a QPC. Electrons can be exchanged between reservoirs $1$ and $2$ across the DQD, and between reservoirs $3$ and $4$ of the QPC. There is no electron transfer between the two channels, but the electrons tunneling through the QPC are affected by the charge state of the quantum dots via Coulomb interaction (wavy lines).}
\label{schematicpicture}
\end{figure}

The DQD channel is made of two quantum dots A and B, each connected to its own reservoir, labeled by $j=1$ and $2$ respectively. The QPC is the junction between reservoirs $j=3$ and $4$. The  circuit is drawn in Fig. \ref{schematicpicture}.

The Hamiltonian of the circuit is given by
\be
H = H_{\rm DQD} + \sum_{j=1}^{4} H_j +V,
\ee
where $H_{\rm DQD}$ denotes the DQD Hamiltonian, 
\be \label{res6}
H_j= \sum_{k} \epsilon_{j k} c_{jk}^{\dagger} c_{jk}
\ee
is the reservoir $j$ Hamiltonian expressed in terms of the creation (anihilation) operators $c_{jk}^{\dagger}$ ($c_{jk}$) of the reservoir single-particle states with energy $ \epsilon_{j k}$, and $V$ is the interaction Hamiltonian between the DQD and the reservoirs.

The DQD Hamiltonian is given by
\be  \label{localbasisH}
H_{\rm DQD}  =  \epsilon_A c_{A}^{\dagger} c_A + \epsilon_B c_{B}^{\dagger} c_B + T \left( c_{A}^{\dagger} c_B + c_{B}^{\dagger} c_A \right)  =  \sum_{s} E_{s} |s \rangle \langle s | ,
\ee
expressed in terms of the single-dot annihilation (creation) operators in each QD, $c_{A/B} $ ($c^{\dagger}_{A/B} $), of single dot states with energies $\epsilon_{A/B}$, and of the tunneling amplitude $T$ between the two dots. This Hamiltonian can alternatively be expressed in terms of the many-body eigenstates $| s \rangle$ of $H_{\rm DQD}$ with energies $E_s$. In the following, we consider a regime in which the sum in the last term of (\ref{localbasisH}) can be restricted to the empty eigenstate $|0 \rangle$ and the single-occupied eigenstates $|+ \rangle$ and $|-\rangle$ (see the Appendix). This assumption is often justified at low temperature since multiple charging in one of the quantum dots requires large amounts of energy.

The interaction Hamiltonian can be split into
\be \label{interactionhamiltonian}
V=  \sum_{j=1}^{2} V_j  +V_{34} + V_{34}^{\rm DQD}.
\ee
The first term is the sum of the tunneling Hamiltonians between the DQD and the reservoirs
\be \fl
V_j =  \sum_{d=A,B}  \sum_{ k}  t_{jd}^{k} \left( c^{\dagger}_{d} c_{jk} + c^{\dagger}_{jk} c_{d} \right)  
 = \sum_{s=+,-} \sum_{ k}  T_{js}^{k} \left( |s \rangle \langle 0 | \, c_{jk} + c^{\dagger}_{jk} \, |0 \rangle \langle s | \right) \label{tunnelinginteraction}
\ee
with the tunneling amplitudes $t_{jd}^{k}$ and $T_{js}^{k} $ in the DQD single and many-body basis, respectively, and for $j=1$ and $2$. Each dot is only connected to its own reservoirs, i.e. $t_{1B}^{k} = t_{2A}^{k} =0$. This part of the Hamiltonian is responsible for the charging and discharging of the DQD through the exchange of electrons with reservoir $j=1$ and $2$.

The QPC reservoirs $3$ and $4$ are directly coupled through the tunneling Hamiltonian
\be
V_{34} = \sum_{kk'} T^{kk'}   \left( c_{3 k}^{\dagger} c_{4 k'} + c_{4k'}^{\dagger} c_{3k}  \right),
\ee
where $ T^{kk'}  $ denote the bare tunneling amplitudes between the QPC reservoirs, i.e. independently of the DQD state.

Finally, electron tunneling in the QPC is also sensible to the charge state of the DQD due to the Coulomb interaction. This interaction is modelled by the capacitive coupling
\bea
V_{34}^{DQD} &  = &   \sum_{kk'} \left( t_{A}^{kk'} c^{\dagger}_{A} c_A +  t_{B}^{kk'} c^{\dagger}_{B} c_B  \right)  \left( c_{3 k}^{\dagger} c_{4 k'} +     c_{4k'}^{\dagger} c_{3k} \right)  \\
& = & \sum_{s,s' } \sum_{kk'} \left(  T_{ss'}^{kk'}  |s \rangle \langle s' | \right)   \left( c_{3 k}^{\dagger} c_{4 k'}+     c_{4k'}^{\dagger} c_{3k}  \right) \label{intQPCDQD7}
\eea
in terms of the capacitive couplings $ t_{A/B}^{kk'}$ and the tunneling amplitudes $T_{ss'}^{kk'} $. Expression (\ref{intQPCDQD7}) shows that some electrons in the QPC can induce transitions in the DQD, exchanging energy with the DQD while tunneling between reservoirs $3$ and $4$.

The Hamiltonian term describing the bare tunneling in the QPC, $V_{34}$, is included in the interaction $V$ which will be subsequently treated to second order in perturbation theory. This is in contrast to our previous work \cite{Cuetara_2013_PhysicalReviewB} where the bare tunneling in the QPC was treated non-perturbatively. The present approach has the advantage to treat all the energy and matter transfers on the same footing, allowing us to develop a consistent thermodynamic description of the full circuit and to evaluate the joint full counting statistics (FCS) of the energy and matter currents in both DQD and QPC channels.


\section{Counting statistics} \label{countingstatistics}





The FCS of the fluxes in weakly coupled open quantum systems can be calculated using the modified quantum master equation formalism \cite{Esposito_2009_ReviewsofModernPhysics}. The statistical properties of the energy and matter currents flowing out of the reservoirs and integrated over a time $t$, $\Delta E_j$ and $\Delta N_j$, are determined by the generating function (GF)
\be \label{stat1}
G(\xi_j , \lambda_j ,t)  = \langle \mbox{e}^{- \sum_{j}\left( \xi_j \Delta E_j + \lambda_j \Delta N_j \right)} \rangle_t  .
\ee
The probability distribution of the fluctuating energy and matter fluxes $J_E^j \equiv - \Delta E_j /t$ and $J_N^j \equiv - \Delta N_j /t$ is then obtained by applying an inverse Fourier transform to the GF (\ref{stat1})
\be \fl \label{probacurr}
P(J_{E}^j , J_{N}^j , t) = \int_{-\infty}^{\infty}\left[ \prod_j t \frac{d \xi_j}{2 \pi} \right]   \int_{0}^{2 \pi} \left[ \prod_j t \frac{d \lambda_j}{2 \pi} \right]  \, e^{i \sum_{j=1}^{4}\left( \xi_j \Delta E_j + \lambda_j \Delta N_j \right)} G(-i \xi_j, -i \lambda_j , t).
\ee

In the following, we evaluate the GF (\ref{stat1}) by performing the FCS of the energy and particle number operators within each reservoir, respectively given by (\ref{res6}) and $N_{j} \equiv \sum_{k} c_{jk}^{\dagger} c_{jk}$. Following \cite{Esposito_2009_ReviewsofModernPhysics}, we introduce the modified Hamiltonian
\be
H ( \xi_j , \lambda_j ) \equiv \mbox{e}^{i \sum_{j=1}^{4} \left( \xi_j H_j +\lambda_j N_j \right)} \, H \,\mbox{e}^{-i \sum_{j=1}^{4} \left(\xi_j H_j + \lambda_j N_j \right)}
\ee
where the counting parameters $\xi_j$ and $\lambda_j$ for $j=1, \dots, 4$ keep track of, respectively, the energy and matter fluctuations in the reservoirs. The DQD and reservoir Hamiltonians, $H_{\rm DQD}$ and $H_j$ respectively, remain unchanged after this transformation. However, the interaction Hamiltonians transform according to
\be 
 V_{j} ( \xi_j , \lambda_j ) = \sum_{s=+,-} \sum_{k} T_{js}^{k}  \left( \mbox{e}^{-i \left(  \xi_{j} \epsilon_{k}+ \lambda_{j} \right) }\, |s \rangle \langle 0 | \, c_{jk} + \mbox{h.c.}\right)  ,
\ee
\bea 
V_{34}( \xi_j , \lambda_j )   =  
  \sum_{kk'}  T^{kk'} \left( \mbox{e}^{ i  \left(  \xi_{3} \epsilon_{k} + \lambda_{3} \right)}\mbox{e}^{-i  \left(  \xi_{4} \epsilon_{k'} + \lambda_{4} \right)} c_{3 k}^{\dagger} c_{4 k'} +   \mbox{h.c.}\right),
\eea
and
\bea  \fl
V^{DQD}_{34}( \xi_j , \lambda_j )   =  \sum_{ss'} \sum_{kk'}  \left( T^{kk'}_{ss'}  |s \rangle \langle s' | \right) 
\left( \mbox{e}^{ i  \left(  \xi_{3} \epsilon_{k} + \lambda_{3} \right)}\mbox{e}^{-i  \left(  \xi_{4} \epsilon_{k'}+ \lambda_{4} \right)} c_{3 k}^{\dagger} c_{4 k'} + \mbox{h.c.} \right).
\eea

With these definitions, the GF can be expressed as 
\be \label{GF1}
G(\xi_j ,\lambda_j ,t) = \mbox{Tr} \left\{ \rho ( i \xi_j , i \lambda_j ,t) \right\} ,
\ee
where the modified density matrix $\rho (\xi_j ,\lambda_j ,t)$ satisfies the modified quantum master equation
\be \label{modifieddynamics}
i \partial_t \rho (\xi_j ,\lambda_j ,t) = H (\xi_j /2 ,\lambda_j /2 ) \rho (\xi_j ,\lambda_j ,t) -\rho (\xi_j ,\lambda_j ,t)  H (-\xi_j /2 ,-\lambda_j /2).
\ee

The initial density matrix $\rho (0)$ of the total system is assumed of the factorized form $\rho(0)= \rho_{\rm S} (0) \prod_{j} \otimes \, \rho_j$, where $\rho_{\rm S} (0)$ is an arbitrary DQD density operator and $\rho_j = \, \exp{\{-\beta_j (H_j-\mu_{j} N_j - \phi_{j})\}}$ denotes the grand-canonical density operator in the reservoir $j$ with inverse temperature $\beta_j=(k_{\rm B}T_j)^{-1}$ and chemical potential $\mu_j$ with $j =1, \dots, 4 $. The corresponding thermodynamic grand-potential is denoted $\phi_j=-\beta^{-1}_j \ln \left[ \mbox{Tr} \left\{ \exp{\{-\beta_j (H_j-\mu_{j} N_j )\}} \right\} \right]$.
This factorization assumption has no implication because only steady state properties will be considered in the following.

In the weak coupling limit to the reservoirs, where the interaction parameters $T_{js}^{k} $, $T^{kk'}$ and $T_{ss'}^{kk'}$ are assumed small enough, the dynamical equation (\ref{modifieddynamics}) leads to a closed modified quantum master equation for the reduced density matrix of the system \cite{Suarez_1992_TheJournalofchemicalphysics, Pechukas_1994_PhysRevLett, Kohen_1997_TheJournalofchemicalphysics, Cheng_2005_TheJournalofPhysicalChemistryB, Jordan_2008_PhysicalReviewA}
\be
\rho_S (\xi_j , \lambda_j , t) = \mbox{Tr}_{ \rm R} \left\{ \rho  (\xi_j , \lambda_j ,t ) \right\} ,
\ee
where $\mbox{Tr}_{\rm R}  \left\{ \cdot \right\}$ denotes a trace over the reservoirs Hilbert space. $\rho_S (0,0 , t)$ is the DQD reduced density matrix.

A common assumption in the present context is that the DQD free oscillations, characterized by the frequencies $\omega_{ss'} = E_s - E_{s'}$, are fast compared to the relaxation time scale $\tau_R$ induced by the reservoirs on the DQD. One can then apply the rotating wave approximation (RWA) \cite{Petruccione_Thetheoryofopenquantumsystems, Gardiner_Quantumnoise, Cohen-Tannoudji1988, Schaller_2008_PhysicalReviewA, VanKampen_Stochasticprocessesinphysicsandchemistry} which consists in an average of the system free oscillations over a time scale $\Delta t$ which is intermediate between
\be
\tau_C \ll \Delta t \ll \tau_R,
\ee
where $\tau_C$ denotes the short correlation time in the reservoirs. As a result, the effective dynamics of the DQD populations, $g_{s}( \xi_j , \lambda_j ,t) = \langle s | \rho_S (i \xi_j , i \lambda_j , t) |s \rangle$, and the coherences, $\langle s | \rho_S (i \xi_j , i \lambda_j , t) |s' \rangle$ for $s \neq s'$, decouple. 




\begin{figure*}[htbp]
\centerline{\includegraphics[width=13cm]{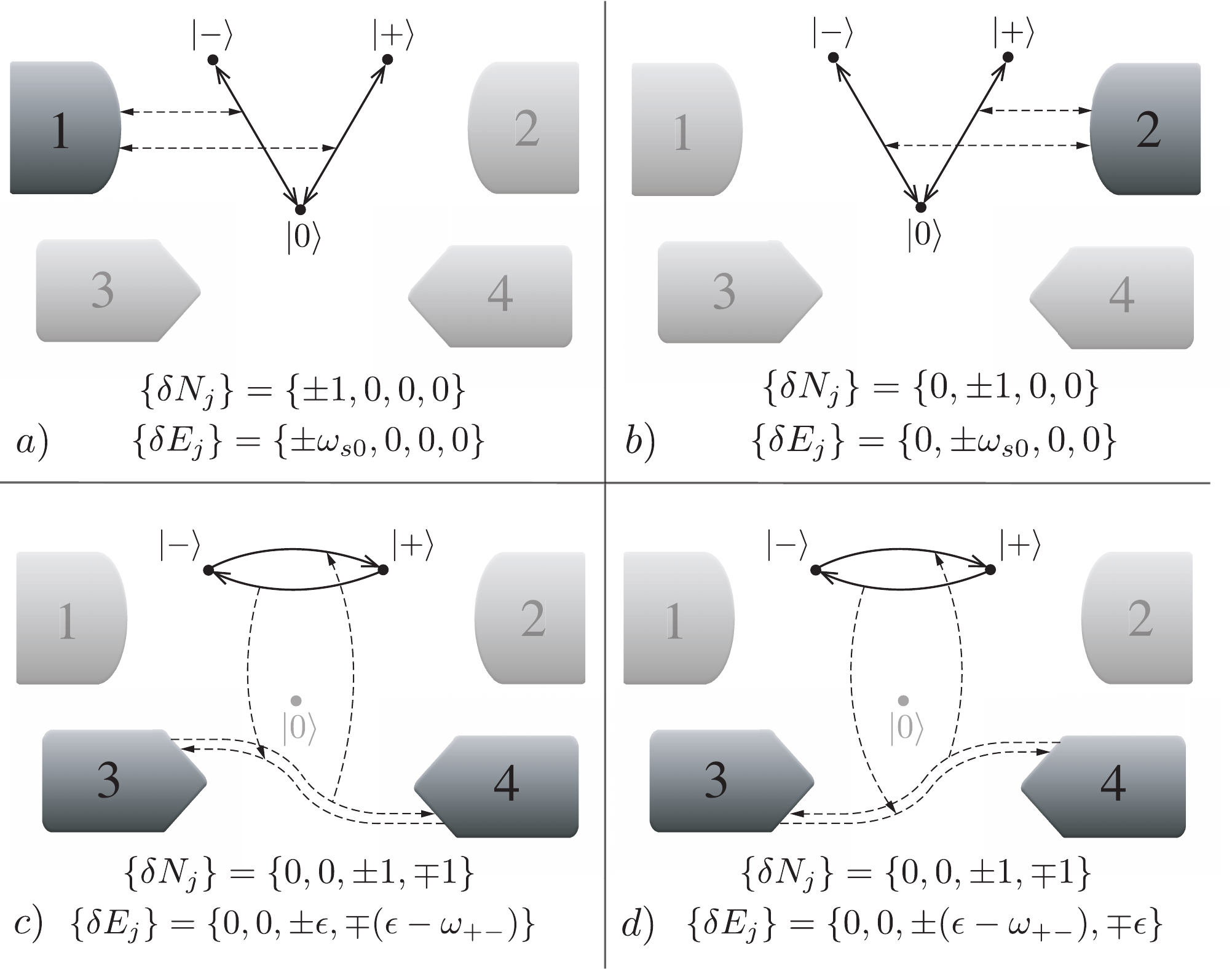}}
\caption{Illustration of the several microscopic processes inducing transitions in the DQD. Each sub-figure illustrates pairs of processes which are time-reversed of each other. The vectors $\left\{ \delta E_j \right\}$ and $\left\{ \delta N_j \right\}$ denote, respectively, the energy and particle number changes in the reservoirs associated to each microscopic processes.}
 \label{DQDtrans}
\end{figure*}

Under the aforementioned hypotheses, the diagonal elements of the DQD reduced density matrix satisfy a Markovian master equation of the form
\be \label{stochmeq}
\partial_t {\bf g}( \xi_j , \lambda_j ,t)=  {\bf W} ( \xi_j , \lambda_j ) \cdot  {\bf g}( \xi_j , \lambda_j ,t) ,
\ee
where we introduced the vector notation
\be \label{modpopvect}
{\bf g}( \xi_j , \lambda_j ,t)=
\left(
\begin{array}{c}
g_{0}(\xi_j , \lambda_j,t) \\
g_{+}(\xi_j , \lambda_j,t) \\
g_{-}(\xi_j , \lambda_j,t) 
\end{array}
\right),
\ee
with the matrix product denoted by '$\cdot$', and where the counting parameters dependent rate matrix ${\bf W} ( \xi_j , \lambda_j ) $ is expressed here below in terms of the transition rates between DQD states.

The reservoirs $j=1$ and $2$ induce random charging and discharging of the DQD due to the tunneling interaction (\ref{tunnelinginteraction}). The corresponding microscopic processes are depicted on Figs. \ref{DQDtrans} a) and b). During such tunneling events, the particle number in reservoir $j=1$ or $2$ changes by an amount $\delta N_j = 1$ ($\delta N_j =-1$) when it charges (discharges) the DQD. On the other hand, the energy change in the reservoir can take the values $\delta E_j = \pm \omega_{s0}$, depending on which many-body state $|s\rangle$, with $s= +$ or $-$, is involved in the transition. The charging and discharging rates induced by reservoirs $j=1$ and $2$ are given by
\be
a_{js}  =  \Gamma_{js}   f_{j}(\omega_{s0})  \\ \mbox{and}  \\ b_{js} =  \Gamma_{js}  (1 -  f_{j}(\omega_{s0}) ) \label{chargedischarge} 
\ee
for $s=+$ or $-$, in terms of the Fermi distribution of single particle states in the reservoir $j$, $f_{j}(x) = (1+\exp \beta_{j}(x-\mu_{j}))^{-1}$, and of the rate constants
\be
 \Gamma_{js}  =  \frac{2 \pi}{\hbar^{2}} \sum_{k} \delta (\epsilon_{jk} -\omega_{s0})  | T^{k}_{j s} |^{2} =  \frac{2 \pi}{\hbar^{2}} \rho_{j}(\omega_{s0}) | T_{js}(\omega_{s0}) |^{2} .
\ee
We took the continuum approximation for the electron density of states in the reservoirs, denoting the energy-resolved tunneling amplitudes by $T_{js}(\epsilon)$, and the density of electron states by $\rho_{j}(\epsilon)$.

On the other hand, the QPC also induces transitions between the DQD states. Though there is no exchange of electrons between the QPC and the DQD, electrons tunneling between the reservoirs $j=3$ and $4$ may exchange energy with the DQD (mainly through photon exchange \cite{Gustavsson_2007_PhysicalReviewLetters}), thus driving transitions between the single-charged states $|+ \rangle$ and $|- \rangle$. These processes are illustrated in Figs. \ref{DQDtrans} c) and d). The corresponding transition rates are given by
\bea 
c_{jj'} (\epsilon) =\Gamma_{jj'} (\epsilon)  \, f_{j} (\epsilon) \, (1-f_{j'} (\epsilon - \omega_{+-})) \nonumber \\
d_{jj'} (\epsilon) = \Gamma_{jj'} (\epsilon)  \, (1- f_{j}(\epsilon)) \, f_{j'} (\epsilon - \omega_{+-}) \label{QPCrates6}
\eea
for $jj'= 34$ and $43$ where the energy-dependent rate constants $ \Gamma_{jj'}(\epsilon)$ are given by
\bea
\Gamma_{34} (\epsilon) & \equiv & \frac{4 \pi}{\hbar^{2}}  \sum_{kk'} |T^{kk'}_{-+} | ^{2} \delta(\epsilon-\epsilon_{3k}) \delta(\epsilon - \omega_{+-} -\epsilon_{4k'}) \\
&  = &  \frac{4 \pi}{\hbar^{2}} \, |T_{-+} (\epsilon, \epsilon  - \omega_{+-})|^{2} \rho_{3}(\epsilon ) \rho_{4} (\epsilon-\omega_{+-} )  \\
\Gamma_{43} (\epsilon) & \equiv & \frac{4 \pi}{\hbar^{2}} \sum_{kk'} |T^{kk'}_{+-} | ^{2} \delta(\epsilon - \omega_{+-}-\epsilon_{3k}) \delta(\epsilon-\epsilon_{4k'}) \\
&  =  & \frac{4 \pi}{\hbar^{2}} \, |T_{+-} (\epsilon - \omega_{+-}, \epsilon )|^{2} \rho_{3}(\epsilon - \omega_{+-}) \rho_{4} (\epsilon )  \\
\eea
in terms of the energy resolved tunneling amplitudes $T_{ss'}(\epsilon, \epsilon')$. Interestingly, the transition rates (\ref{QPCrates6}) are written as a product of Fermi functions in both QPC reservoirs and as such, cannot be written as a sum of individual reservoir contributions. 

The tunneling events between the DQD and the reservoirs $j=1$ and $2$ contribute to the rate matrix through the matrix elements
\bea 
\left[ {\bf W} ( \xi_j , \lambda_j ) \right]_{s0}  \equiv  \sum_{j=1,2} a_{js} \, \mbox{e}^{ \left( \xi_{j} \omega_{s0} + \lambda_j \right)} \label{DQDrates16} \\ 
\left[ {\bf W}( \xi_j , \lambda_j ) \right]_{0s} \equiv  \sum_{j=1,2}  b_{js} \, \mbox{e}^{-\left( \xi_{j} \omega_{s0} +\lambda_j \right)}   \label{DQDrates26},
\eea
where the counting parameters $\xi_j$ and $\lambda_j$ keep track of the net fluxes of energy $\pm \omega_{s0}$ and particles $\pm 1$ flowing out of reservoir $j$ at each such transition.

The contribution from the QPC transfers can be separated into two categories, depending on whether or not the electrons tunneling between reservoirs $3$ and $4$ exchange energy with the DQD. In the first case, tunneling events in the QPC contribute to the rate matrix through the components
\bea 
 \left[ {\bf W} ( \xi_j , \lambda_j ) \right]_{+-}   = \sum_{jj'} \int d\epsilon \,   c_{jj'} (\epsilon) \, \mbox{e}^{ \left( \xi_j \epsilon -\xi_{j'} (\epsilon - \omega_{+-}) + \lambda_{j} - \lambda_{j'} \right)}  \label{QPCrate16}
  \\
 \left[ {\bf W} ( \xi_j , \lambda_j ) \right]_{-+}   = \sum_{jj'} \int d\epsilon  \, d_{jj'} (\epsilon) \, \mbox{e}^{ \left(  -\xi_j \epsilon   +\xi_{j'} (\epsilon- \omega_{+-}) - \lambda_j + \lambda_{j'}  \right) }  \label{QPCrate26}
\eea
where the sum in these last two equalities runs over $jj'=34$ and $43$. Each transition involves a net transfer of one electron from reservoir $j$ to reservoir $j'$ of the QPC or vice versa. If the energy of the outgoing electron coming from $j$ is $\epsilon$, it enters reservoir $j'$ with energy $\epsilon \pm \omega_{+-}$ depending on whether it emits or absorbs energy from the DQD.

\begin{figure*}[htbp]
\centerline{\includegraphics[width=6.5cm]{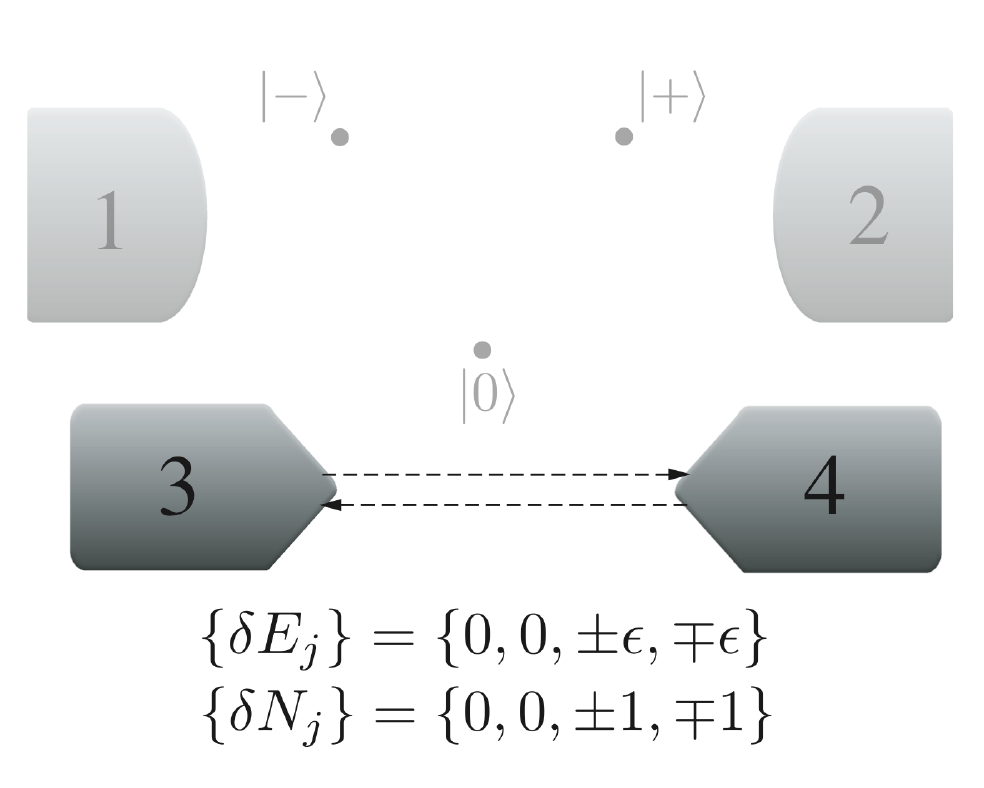}}
\caption{Illustration of the tunneling processes in the QPC that do note induce transitions in the DQD.}
 \label{transitionsQPC}
\end{figure*}

Finally, the contributions from the electrons tunneling in the QPC without exchanging energy with the DQD appear along the diagonal elements of the rate matrix
\be \label{statedepCGF}
\left[ {\bf W} ( \xi_j , \lambda_j ) \right]_{ss}  = - \sum_{s'\neq s }  \left[ {\bf W} ( 0 , 0 ) \right]_{s's} + G_{s} ( \xi_j , \lambda_j ) .
\ee
The first term in the right-hand side of this equation ensures the conservation of the probability for the occupation probabilities in the DQD when the counting parameters are set to zero while the second one accounts for the tunneling of electrons through the QPC without interaction with the DQD (see Fig. \ref{transitionsQPC}). As a matter of fact, $G_{s} ( \xi_j , \lambda_j ) $ is the GF of the energy and particle transfer in the QPC given that the DQD is in state $|s\rangle$, i.e.
\bea \nonumber \fl
G_{s}  ( \xi_j , \lambda_j )  
= \int \, d\epsilon \, \gamma_{s}(\epsilon)  \\ \fl
\left[ f_{3}(\epsilon) (1-f_{4} (\epsilon)) \left(1- e^{  (\lambda_{3} - \lambda_{4})}e^{ \epsilon (\xi_{3} - \xi_{4})} \right) \right. 
\left.+  f_{4}(\epsilon) (1-f_{3} (\epsilon)) \left(1- e^{-(\lambda_{3} - \lambda_{4})}e^{- \epsilon (\xi_{3} - \xi_{4})} \right)  \right] . \label{diagorate}
\eea
It turns out that this is the Levitov-Lesovik formula \cite{Levitov_1993_JETPLETTERSC, Levitov_1996_JournalofMathematicalPhysics, Schoenhammer_2007_PhysicalReviewB, Klich_2002_eprintarXiv, Gogolin_2006_PhysicalReviewB} to second order in the tunneling amplitude $T_{s}(\epsilon)$, or equivalently, to first order in
\be \label{gammatunneling}
 \gamma_{s} (\epsilon) \equiv 2\pi |V_{ss}(\epsilon)|^{2} \rho_{3} (\epsilon) \rho_{4} (\epsilon),
\ee
which results from the fact that we treated the interaction (\ref{intQPCDQD7}) perturbatively. In contrast to many previous work on transfer through quantum dots (such as \cite{Cuetara_2011_PhysicalReviewB}), where each microscopic process is associated to an actual transition in the QD, the present circuit provides a nice example of an open quantum system in which the microscopic processes do not necessarily affect the system (DQD) populations while contributing to the energy and particle flows out of the reservoirs. The modified quantum master equation formalism is thus essential in order to keep track of these processes.

The Fourier transform of the diagonal matrix elements (\ref{modpopvect})
\be \fl
{\bf p} (\Delta E_j , \Delta N_j , t) 
=\int_{-\infty}^{\infty}\left[ \prod_j \frac{d \xi_j}{2 \pi} \right]   \int_{0}^{2 \pi} \left[ \prod_j \frac{d \lambda_j}{2 \pi} \right]  \, e^{ - i \sum_{j=1}^{4}\left( \xi_j \Delta E_j + \lambda_j \Delta N_j \right)}\, {\bf g} (-i \xi_j , -i \lambda_j  , t) 
\ee
gives the join probability distribution $\left[ {\bf p}(\Delta E_j , \Delta N_j, t)  \right]_{s} = p_s(\Delta E_j , \Delta N_j , t)$ of observing the system in state $|s \rangle $ at time $t$ and the changes in energies $\Delta E_j$ and particle numbers  $\Delta N_j$ in each reservoir.

By applying a Fourier transform to the modified rate equation (\ref{stochmeq}), we get a master equation describing the dynamics of the DQD as well as the exchange processes with the reservoirs
\be \label{rateeq6} \fl
\partial_t {{\bf p}} (\Delta E_j , \Delta N_j , t) = \int \prod_j \left[ d\delta E_j \right]\left[ d\delta N_j \right]  \hat{{\bf W}} (\delta E_j , \delta N_j ) \cdot {\bf p}(\Delta E_j - \delta E_j, \Delta N_j -\delta N_j, t) ,
\ee
where the rate matrix $\hat{{\bf W}} (\delta E_j , \delta N_j )$ is obtained as the Fourier transform of the modified rate matrix
\be \label{ratematenpart} \fl
\hat{{\bf W}} (\delta E_j , \delta N_j ) =\int_{-\infty}^{\infty}\left[ \prod_j \frac{d \xi_j}{2 \pi} \right]   \int_{0}^{2 \pi} \left[ \prod_j \frac{d \lambda_j}{2 \pi} \right] \, e^{ - i \sum_{j} \left( \xi_j \delta E_j + \lambda_j \delta N_j \right)}  \, {\bf W} (-i \xi_j ,  -i\lambda_j ).
\ee
The Fourier transform of the modified rate matrix is easily taken by using the relations
\be
\int^{2\pi}_{0} \frac{dx}{2 \pi} \mbox{e}^{i x \alpha}  =  \delta_{\alpha ,0} \\ \mbox{and} \\
\int_{-\infty}^{\infty} \frac{dx}{2 \pi} \mbox{e}^{i x \alpha}  =  \delta (\alpha )
\ee
in terms of the Kronecker delta symbol $\delta_{\alpha ,0}$ and the Dirac delta distribution $\delta (\alpha)$. Accordingly, the rate matrix $\hat{{\bf W}} (\delta E_j , \delta N_j ) $ is obtained by making the following substitutions
\be
\mbox{e}^{\pm  \xi_j \alpha_j}  \rightarrow  \delta (\delta E_j \mp \alpha_j) \nonumber \\ \mbox{and} \\
 \mbox{e}^{\pm  \lambda_j }  \rightarrow  \delta_{\delta N_j , \mp 1}
\ee
in the modified rate matrix elements (\ref{DQDrates16})-(\ref{statedepCGF}).

By integrating equation (\ref{rateeq6}) over the energy and particle fluctuations $\Delta E_j$ and $\Delta N_j$, or equivalently by setting the counting parameters to zero in the modified rate equation (\ref{stochmeq}), we obtain a stochastic master equation for the occupation probabilities in the DQD
\be 
\dot{{{\bf p}}}(t) = {\bf W} \cdot {\bf p}( t) 
\ee
with rate matrix given by
\be \label{ratemat6} \fl
 {\bf W} = \left( \begin{array}{ccc}
-a_{1+}-a_{2+} - a_{1-} - a_{ 2-} & b_{1+}+b_{2+} & b_{1-}+b_{2-} \\
a_{1+}+a_{2+}& -b_{1+}-b_{2+} -d_{34} - d_{43}& c_{34}+c_{43}\\
 a_{1-} + a_{ 2-} & d_{34} + d_{43}& -c_{34}-c_{43} -b_{1-}-b_{2-} \\
\end{array} \right),
\ee
and where
\be \nonumber
c_{jj'} = \int d\epsilon \,   c_{jj'} (\epsilon) \\ \mbox{and} \\
d_{jj'} = \int d\epsilon \,   d_{jj'} (\epsilon).  \label{intratesQPC}
\ee

Now, by formally solving the rate equation (\ref{stochmeq}), the GF of the energy and matter currents can be expressed as
\be \label{GFsol}
G(\xi_j , \lambda_j ,t) = {\bf 1} \cdot \mbox{e}^{{\bf W} ( \xi_j , \lambda_j ) t} \cdot {\bf p}_{0} ,
\ee
where ${\bf 1}$ denotes the line vector $(1 , 1 , 1)$ while ${\bf p}_{0}$ is the initial occupation probability of the DQD states. All the moments of the currents can be obtained by taking multiple derivatives of the GF with respect to the counting parameters.

At steady state, the statistics of the currents is captured by the cumulant generating function (CGF)
\bea \label{CGF6}
\mathcal{G} (\xi_j, \lambda_{j}) & \equiv & -\lim_{t \rightarrow \infty} \frac{1}{t} \ln G (\xi_j , \lambda_j , t) .
\eea
Using expression (\ref{GFsol}) for the GF, one sees that the CGF is obtained as the dominant eigenvalue of the rate matrix ${\bf W} ( \xi_j , \lambda_j )$, independently of the initial condition on the system $ {\bf p}_{0}$.


\section{Nonequilibrium thermodynamics} \label{noneqthermodynamics}



\subsection{Local detailed balance and fluctuation theorem} \label{detailedbalancesection}


The charging and discharging rates (\ref{chargedischarge}) depicted on Figs. \ref{DQDtrans} a) and b) satisfy the LDB condition \cite{Kubo_Statistical-mechanicaltheory}
\be \label{kms6}
\ln \frac{a_{js}}{b_{js}} =-\beta_j (\omega_{s0} - \mu_j)
\ee
in terms of the inverse temperature $\beta_j$ and chemical potential $\mu_j$ of the reservoir involved in the transition. This property is  a direct consequence of the Kubo-Martin-Schwinger (KMS) condition which is satisfied by the equilibrium correlation functions of the reservoirs \cite{Esposito_2009_ReviewsofModernPhysics}.

Regarding the QPC induced transitions on the DQD, it has previously been noted \cite{Cuetara_2013_PhysicalReviewB, Schaller_OpenQuantumSystemsFarfromEquilibrium} that the corresponding total rates do not satisfy a LDB, unless the QPC is assumed to be at equilibrium ($\beta_3 = \beta_4$ and $\mu_3 = \mu_4$). However, we showed in the previous section that in the weak coupling limit, one can identify the pairs of microscopic processes related by time reversal (cf. Figs. \ref{DQDtrans} c) and d)), and write the total rates as a sum of contributions from such elementary processes, see equations (\ref{QPCrate16}) and (\ref{QPCrate26}) together with (\ref{QPCrates6}). Each of these contributions satisfies the LDB condition
\be \label{kms62}
\ln \frac{c_{jj'} (\epsilon)}{d_{jj'} (\epsilon)} = -\beta_j (\epsilon - \mu_j)+\beta_{j'} (\epsilon-\omega_{+-} - \mu_{j'}),
\ee
where $c_{jj'}(\epsilon)$ is the rate at which electrons with energy $\epsilon$ tunnel from reservoir $j$ to $j'$ in the QPC while releasing an amount of energy $\omega_{+-}$ to the DQD, and $d_{jj'} (\epsilon)$ is the rate of the associated time-reversed process.

We note that the right-hand sides of expressions (\ref{kms6}) and (\ref{kms62}) is nothing but the entropy flowing from the reservoirs during these processes
\be \label{heatflow}
\Delta_{ss'} S (\delta E_j , \delta N_j )  = - \sum_j \beta_j Q_j  \mbox{, with } Q_j = -\delta E_{j} + \mu_j \delta N_j
\ee
and where the components of the vectors
\be \fl
\left\{  \delta E_j \right\}  =  \left\{ \delta E_1 , \delta E_2 , \delta E_3 , \delta E_4 \right\} \\ \mbox{and} \\
\left\{  \delta N_j \right\}  =  \left\{ \delta N_1 , \delta N_2 , \delta N_3 , \delta N_4 \right\}
\ee
denote the changes in energy and particle number in the reservoirs associated to each microscopic transition, as given in Fig. \ref{DQDtrans}.  Relations (\ref{kms6}) and (\ref{kms62}) can then both be rewritten in terms of the rate matrix (\ref{ratematenpart}) as
\be \label{genKMS}
\ln \frac{ \left[ \hat {\bf W} (\delta E_j , \delta N_j) \right]_{ss'} }{\left[ \hat {\bf W}  (-\delta E_j ,- \delta N_j) \right]_{s's} } =  \Delta_{ss'} S (\delta E_j , \delta N_j ) ,
\ee
where the transition rate $\left[\hat  {\bf W}  (-\delta E_j ,- \delta N_j) \right]_{s's}$ corresponds to the time-reversed process of the one associated to $\left[ \hat {\bf W} (\delta E_j , \delta N_j) \right]_{ss'}$. 
Relation (\ref{genKMS}) implies that the modified transition rates (\ref{DQDrates16})-(\ref{statedepCGF}) satisfy
\be
\left[{\bf W} (\xi_j , \lambda_j)  \right]_{ss'}= \left[{\bf W} (\beta_j - \xi_j , -\beta_j \mu_j - \lambda_j)  \right]_{s's}.
\ee
These relations also hold for the diagonal elements of the rate matrix, given in equation (\ref{statedepCGF}), $G_s (\xi_j , \lambda_j) = G_s (\beta_j - \xi_j , - \beta_j \mu_j - \lambda_j) $.

They ensure the invariance of the characteristic polynomial of the matrix ${\bf W} (\xi_j , \lambda_j)  $ under the transformations
\be
 \xi_j  \rightarrow  \beta_j - \xi_j \\ \mbox{and} \\
\lambda_j  \rightarrow   -\beta_j \mu_j - \lambda_j .
\ee
Since the CGF is obtained as the largest eigenvalue of the modified rate matrix ${\bf W} ( \xi_j ,  \lambda_j)  $, this invariance property leads to the symmetry
\be \label{sym6}
\mathcal{G} (\xi_j, \lambda_{j}) = \mathcal{G} ( \beta_j -\xi_j, -\beta_j \mu_j - \lambda_{j}).
\ee
We further note that the CGF only depends on the differences
\be \label{differences_of_counting_param}
\xi_1-\xi_4 , \, \xi_2-\xi_4 , \, \xi_3-\xi_4 \\ \mbox{and}  \\ \lambda_1-\lambda_2 ,  \, \lambda_3-\lambda_4 .
\ee
This can be directly shown by verifying that the characteristic polynomial of the rate matrix ${\bf W} ( \xi_j ,  \lambda_j) $ only depends on the differences (\ref{differences_of_counting_param}). It is a direct consequence of the conservation of the total energy and particle number. In addition, the particle numbers in each sub-channel is also conserved since electrons cannot tunnel between the DQD and the QPC. Accordingly, we introduce the CGF
\be \label{gendiffs}
\tilde \mathcal{G} (\xi_l, \lambda_1, \lambda_3) \equiv \left. \mathcal{G} (\xi_j, \lambda_j)  \right|_{\xi_4 = \lambda_2 = \lambda_4 = 0}.
\ee
where $l=1,2,3$. The CGF (\ref{gendiffs}) satisfies the steady state current FT symmetry \cite{Andrieux_2007_JournalofStatisticalPhysics, Esposito_2007_PhysRevE}
\be \label{fluctuationtheorem}
\tilde \mathcal{G} (\xi_l, \lambda_1, \lambda_3)  = \tilde \mathcal{G} (A_{ E}^l  -\xi_l,  A_{ N}^1-\lambda_1 , A_{ N}^3-\lambda_3 )
\ee
in terms of the thermodynamic forces applied to the system:
\be \label{affinities1}
A_{ E}^l \equiv  \beta_4 - \beta_l  \\ \mbox{for} \\ l=1,2,3 
\ee
\be \label{affinities2}
 A_{ N}^1 \equiv  \beta_1 \mu_1 - \beta_2 \mu_2 \\  \mbox{and} \\    A_{ N}^3 \equiv \beta_3 \mu_3 - \beta_4 \mu_4.
\ee

More explicitly, this FT can be expressed in terms of the join probability distribution 
\be
\tilde P (J_{E}^{l} , J_{N}^{1}, J_{N}^{3} ,t ) = \int d J_{E}^{4} \int  d J_{N}^{2} \int d J_{N}^{4} \, P (J_{E}^{j} , J_{N}^{j} ,t )
\ee
as
\be \label{detailedFT}
\lim_{t \rightarrow \infty} \frac{1}{t} \ln \frac{ \tilde P(J_{E}^{l} , J_{N}^{1}, J_{N}^{3} ,t )}{ \tilde P(-J_{E}^{l} , -J_{N}^{1}, -J_{N}^{3} ,t )} =  \sum_{l=1}^{3} A_{ E}^l J_{ E}^l + A_{ N}^1 J_{N}^1+ A_{ N}^3 J_{N}^3 ,
\ee
which makes explicit reference to the steady-state entropy production (right-hand side) generated by the fluxes against the thermodynamic affinities (\ref{affinities1})-(\ref{affinities2}).

In the isothermal setups, one obtains a bivariate FT for the join distribution of particle currents through each channel
\be 
\lim_{t \rightarrow \infty} \frac{1}{t} \ln \frac{ P(J_{N }^1 , J_{N }^3 ,t)}{P(-J_{N }^1 , -J_{N }^3 ,t)} = \beta ( \mu_1 - \mu_2) J_{N}^1 + \beta (\mu_3 - \mu_4)  J_{N}^3,
\ee
thus extending previous result obtained in \cite{Cuetara_2013_PhysicalReviewB}, where the statistics of the current in the QPC was not assessed.


\subsection{Mean currents and entropy production}\label{meancur6}


Using (\ref{CGF6}) together with (\ref{GFsol}), we can formally write the steady-state mean energy and particle currents out of reservoir $j$ as
\bea \label{imatter}
\langle J_{E}^{j} \rangle& \equiv & - \lim_{t \rightarrow \infty} \frac{\langle \Delta E_{j} \rangle }{t} =   {\bf 1} \cdot \partial_{\xi_{j}} {\bf W} ( 0,0) \cdot {\bf P}  \\ \label{ienergy}
\langle J_{N}^{j}  \rangle & \equiv &-  \lim_{t \rightarrow \infty} \frac{ \langle  \Delta N_{j} \rangle }{t} =  {\bf 1} \cdot \partial_{\lambda_{j}} {\bf W} ( 0,0) \cdot {\bf P} ,
\eea
where $ {\bf P}  = \lim_{t\rightarrow \infty} \mbox{e}^{ t {\bf W}}  \cdotp {\bf p }_{0} $
denotes the vector of steady-state occupation probabilities $ P_s = \left[ {\bf P}  \right]_{s} $ on the DQD. The steady-state probabilities are directly obtained by solving the equation ${\bf W}  \cdotp {\bf P } =0 $.

Using the rate matrix of the present model ${\bf W}$ given by (\ref{ratemat6}), we find for  outgoing currents of particles and energy from reservoirs $j=1$ and $2$ that
\be \label{materDQD}
\langle J_{N}^{j}  \rangle   =  \sum_{s=+,-} \left( a_{js} P_{0} - b_{js} P_{s}  \right), \\ 
\langle J_{E}^{j}  \rangle =  \sum_{s=+,-} \omega_{s0} \left( a_{js} P_{0} - b_{js} P_{s}  \right).
\ee
The particle currents for reservoirs $3$ and $4$ can be expressed as the sum of two contributions
\be \label{totcurr3}
\langle J_{N}^{3}\rangle_\nu = \int d\epsilon \, \langle J_{N}(\epsilon) \rangle_\nu , \quad \mbox{for} \quad \nu=d,i  ,
\ee
in terms of the energy resolved currents
\bea  \label{indcurrent}
\langle J_{N} (\epsilon ) \rangle_i= \sum_{s} \gamma_{s}(\epsilon) \left( f_{3} (\epsilon) - f_{4} (\epsilon) \right) \\ \label{curentdep}
\langle J_{N} (\epsilon ) \rangle_d =   (c_{34} (\epsilon) - c_{43}  (\epsilon)) P_{-} + ( d_{43}  (\epsilon) - d_{34} (\epsilon)) P_{+}  .
\eea
The current $\langle J_{N}  \rangle_i$ is conveyed by electrons which tunnel between the QPC reservoirs without affecting the DQD while $\langle J_{N}\rangle_d$ is the current of electrons that induce transitions between states $|+\rangle \rightleftharpoons |- \rangle$ in the DQD through the exchange of an amount $\pm \omega_{+-}$ of energy with the DQD. The energy currents out of the reservoirs $3$ and $4$ are in turn given by
\bea
\langle J_{E}^{3} \rangle& = & \int d \epsilon \, \epsilon \, \left(\langle J_{N}(\epsilon) \rangle_d + \langle J_{N}(\epsilon) \rangle_i \right)+  \omega_{+-} \left( c_{43}  P_{-} - d_{43} P_{+} \right) \label{energyQPC2}  \\
\langle J_{E}^{4} \rangle & = & - \int d\epsilon \, \epsilon \,\left(\langle J_{N}(\epsilon) \rangle_d + \langle J_{N}(\epsilon) \rangle_i \right)+  \omega_{+-} \left( c_{34}  P_{-} - d_{34} P_{+} \right)  \label{energyQPC}.
\eea
From these expressions, one readily verifies the conservation laws
\be
\sum_{j=1}^{4} \langle J_{E}^{j}\rangle  =  0,  \\ \langle J_{N}^{1}\rangle  =  - \langle J_{N}^{2}\rangle  \\ \mbox{and} \\
\langle J_{N}^{3}\rangle  =\langle J_{N}^{3}\rangle_d +\langle J_{N}^{3}\rangle_i   = - \langle J_{N}^{4}\rangle \label{conscurn}.
\ee

The entropy flow from the environment reads
\be \label{entrchangeenv}
\langle J_{S_r} \rangle = -\sum_{j=1}^{4} \beta_{j} (\langle J_{E}^{j} \rangle- \mu_{j} \langle J_{N}^{j} \rangle),
\ee
while the rate of entropy production $\dot{S}_{i}$ in the whole setup can be expressed as
\be
\dot{ S}_{i} = \partial_t { S} + \langle J_{S_r} \rangle,
\ee
where $S = -\sum_s p_s \ln p_s$ denotes the Shannon entropy of the system whose time derivative vanishes at steady state. As a result, the steady-state rate of entropy production can be linked to the energy and matter currents by
\bea \label{rel16}
\dot{S}_{i}  =  \langle J_{S_r} \rangle  =  -\sum_{j=1}^{4} \beta_{j} (\langle J_{E}^{j} \rangle- \mu_{j} \langle J_{N}^{j} \rangle).
\eea

Using the conservation laws (\ref{conscurn}), this equation can be rewritten as a sum of terms that can be interpreted as the dissipation generated by each current against its thermodynamic affinity
\bea 
\dot{S}_{i} =\sum_{l=1}^{3} A_{ E}^l \langle J_{ E}^l \rangle + A_{ N}^1\langle J_{N}^1\rangle + A_{ N}^3\langle J_{N}^3 \rangle . \label{entropprod}
\eea
This expression takes the same form as the left-hand side of equation (\ref{detailedFT}), which can be used to prove that it is always non-negative, i.e. $\dot{S}_{i} \geq 0$. The entropy production (\ref{entropprod}) plays a central role for the thermodynamic analysis of our device driven out of equilibrium by thermal and chemical potential gradients. It is essential in order to define proper notions of efficiency when the device is set to work as a thermodynamic machine. In the next section, we make use of our analysis in order to characterize the performance of our device when operating as a thermoelectric and an isothermal electric converter.


\section{Device operating as a thermodynamic machine} \label{engines}



\subsection{Regime of thermoelectric conversion}\label{maxpowersection}


We consider a regime in which the QPC is the hot reservoir with inverse temperature $\beta_h = \beta_3 = \beta_4$ whereas the DQD reservoirs $j =1$ and $2$ constitute the cold reservoir with inverse temperature $\beta_c = \beta_1 = \beta_2$. A fraction of the heat flow from the QPC
\be
\dot{\mathcal{Q}}  =   J_{E}^{3}  + J_{E}^{4}  \label{inputheat}
\ee
may then be converted by the DQD into electro-chemical work against a bias $\Delta \mu \equiv \mu_1 - \mu_2$ applied between the reservoirs $1$ and $2$
\be
\dot{\mathcal{W}}  = - \Delta \mu  J_{N}^{1}  \label{outputpower} .
\ee
We assume a vanishing bias in the QPC so that the working regime of our heat engine is
\be \label{tempgrad}
\beta_h <\beta_c, \\ \mu_3 = \mu_4 \\ \mbox{and} \\ \Delta \mu  <0 .
\ee
The irreversible entropy production (\ref{entropprod}) reduces to
\be
\dot{S}_{i} = - \beta_c  \langle \dot{\mathcal{W}}  \rangle  + \left( \beta_c -\beta_h \right)  \langle \dot{\mathcal{Q}}  \rangle  \geq 0 ,
\label{entropprodthermalmach}
\ee
where $ \langle \dot{\mathcal{W}}  \rangle$ is the average output power and $\langle \dot{\mathcal{Q}}  \rangle$ the average heat flow from the QPC. One observes that a positive output power always contributes as a negative term to the entropy production, which is compensated by the heat term $ \left( \beta_c -\beta_h \right)  \langle \dot{\mathcal{Q}}  \rangle$ so as to satisfy the second law inequality (\ref{entropprodthermalmach}).

The efficiency $\eta$ of the heat to work conversion process described above is defined by the ratio
\be
\eta \equiv \frac{\langle \dot{\mathcal{W}}\rangle}{\langle \dot{\mathcal{Q}}\rangle } = \frac{ - \Delta \mu \langle J_{N}^{1} \rangle}{\langle J_{E}^{3} \rangle +\langle J_{E}^{4} \rangle } \leq \eta_C \label{effint6} ,
\ee
where $\eta_C=1- \beta_h/ \beta_c $  is the Carnot efficiency of the machine and the inequality is a direct consequence of the positivity of the entropy production (\ref{entropprodthermalmach}). The Carnot efficiency may only be reached for a heat engine working reversibly, i.e. satisfying $\dot S_i =0$. However, such machines work infinitely slowly so that the extracted power vanishes in the limit of a reversible machine. This issue has motivated the study of the efficiency  at maximum output power in thermal engines \cite{Curzon_1975_AmericanJournalofPhysics, VandenBroeck_2005_PhysicalReviewLetters, Tu_2008_JournalofPhysicsA, Schmiedl_2008_EPL, Esposito_2009_EPL, Esposito_2010_Physicalreviewletters, Sanchez_2011_prb}. In particular, the Curzon-Ahlborn efficiency
\be \label{caeff}
\eta_{CA} = 1- \sqrt{\frac{\beta_h}{\beta_c}}
\ee
has been shown to provide a universal upper bound on the efficiency at maximum power in machines working in the linear regime \cite{VandenBroeck_2005_PhysicalReviewLetters}, as well as a good reference in the non-linear regime \cite{Esposito_2010_Physicalreviewletters}. Reaching maximal output power and optimal conversion efficiency requires a fine tuning of the device parameters. For fixed $\Delta \mu$, the DQD spectrum and its coupling to the reservoirs via the tunneling amplitudes need to be adjusted. In particular, highest efficiencies are attained in the so-called regime of tight-coupling where the input and output power become proportional to each other \cite{Esposito_2010_Physicalreviewletters, Sanchez_2011_prb}.

\begin{figure*}[htbp]
\centerline{\includegraphics[width=13cm]{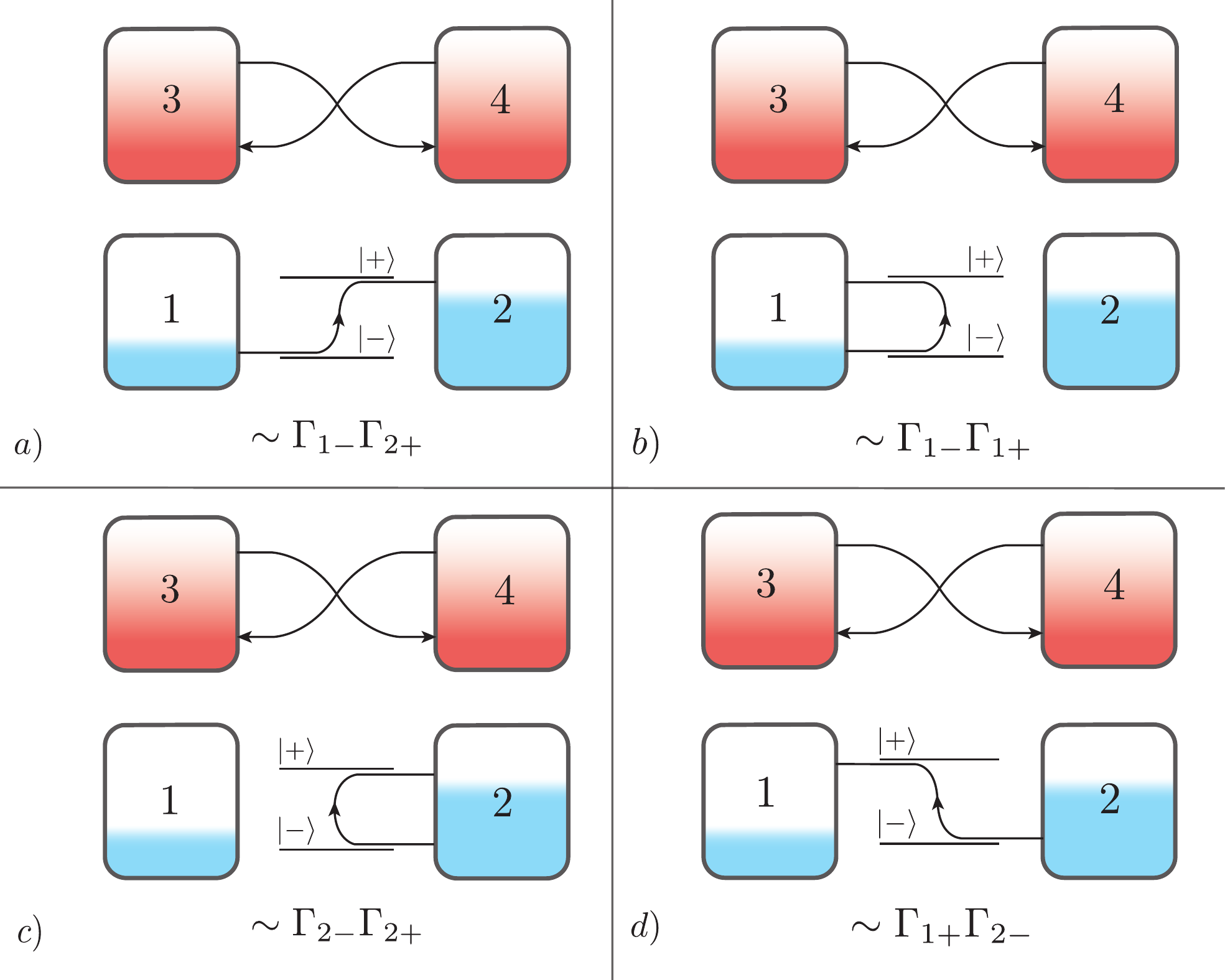}}
\caption{Illustration of the different processes involving a transtion from state $|-\rangle$ to state $|+\rangle$ in the DQD.}
 \label{dragtrans}
\end{figure*}

This can be understood in the present context by comparing the rates of the second order processes depicted in Fig. \ref{dragtrans}. In presence of the temperature gradient (\ref{tempgrad}), the QPC will preferentially give energy to the DQD by inducing transitions from state $|-\rangle$ to state $|+\rangle$. The microscopic process leading to a net flow of charge against the bias $\Delta \mu$ and involving such transitions is depicted in Fig. \ref{dragtrans} a). Without optimization, this process is not more likely to happen than the other processes depicted in Fig. \ref{dragtrans}, which do not involve a charge transfer in the desired direction thus lowering the output power as well as the efficiency of the heat to work conversion. However, provided the tunneling rates satisfy
\be \label{condmach}
\Gamma_{1+}, \Gamma_{2-} \ll \Gamma_{1-},\Gamma_{2+} ,
\ee
the process depicted in \ref{dragtrans} a) becomes the dominant one, leading to a highly efficient conversion of the heat flow into electric output power.

In the tight-coupling limit, i.e. for $\Gamma_{1+}, \Gamma_{2-} \rightarrow 0$, the matter current through the DQD and the energy current from the QPC are totally correlated, their average being thus proportional to each other
 \be
 \langle J_{E}^{3} \rangle +\langle J_{E}^{4} \rangle =  \omega_{+-}\langle J_{N}^{1} \rangle,
\ee
as is verified by using the explicit expressions for the currents given in section \ref{meancur6}. Consequently, the efficiency takes the simple form
\be \label{effmach2}
\eta = \frac{ - \Delta \mu}{\omega_{+-}}.
\ee
This last relation shows that in the tight coupling regime, our device only works as a heat engine producing a positive output power in the range $0<  - \Delta \mu < \omega_{+-}$.
The similarity of expression (\ref{effmach2}) and the one obtained for the efficiency of the machine considered in Ref. \cite{Sanchez_2011_PhysRevB} stems from the quantized character of the amount of energy exchanged between the reservoirs and the work converter in both models. In the present case, the QPC exchanges energy with the DQD in the form of quantas whose energy is given by $\omega_{+-}$.

\begin{figure*}[htbp]
\centerline{\includegraphics[width=15.5cm]{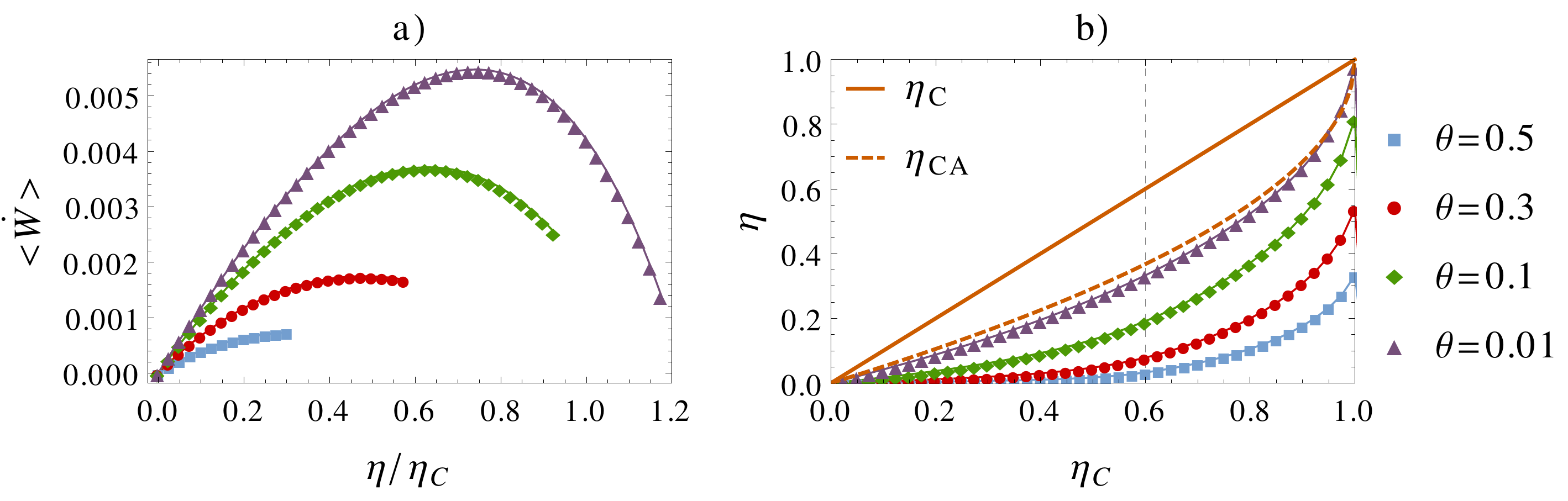}}
\caption{a) Average output power as a function of the rescaled efficiency $\eta / \eta_C$ for different values of the asymmetry parameter $\theta$ defined in (\ref{assympar6}). The Carnot efficiency was taken as $\eta_C = 0.6$. b) Efficiency at maximum power as a function of Carnot efficiency. The continuous red line is for Carnot efficiency while the dotted one follows the Curzon-Ahlborn efficiency (\ref{caeff}). Remaining parameters were chosen for both plots as $\beta_c=1$, $E_0 = 0$, $V_3 = V_4 = 0$, $(\mu_1 + \mu_2)/2 = 0.7$, $\Gamma_{34} = \Gamma_{43}=0.01$, $\Gamma_{1-} = \Gamma_{2+} =0.1$. The energies $E_+$ and $E_-$ of the single-occupied states of the DQD, as well as the bias $\Delta \mu$ in the case of the right plot, were numerically adjusted in order to reach a maximal output power.}
 \label{powerfig}
\end{figure*}

In Fig. \ref{powerfig} a), the average output power is plotted against the rescaled efficiency $\eta/\eta_C$ for different values of the ratio
\be \label{assympar6}
\theta \equiv \frac{\Gamma_{1+}}{\Gamma_{1-}} = \frac{\Gamma_{2-}}{\Gamma_{2+}} ,
\ee
which measures the distance from the tight coupling regime (\ref{condmach}). For each curve, the Carnot efficiency was held at the fixed value $\eta_{C} = 0.6$, whereas the DQD spectrum was numerically adjusted to maximize the output power at fixed $\Delta \mu$. Only values in the range where the device works as a heat engine are shown. As can be seen, the range over which a positive output power can be produced as well as its magnitude decrease as one moves away from the tight coupling regime, i.e. as $\theta$ increases.

Fig. \ref{powerfig} b), shows curves of the efficiency where the DQD spectrum as well as the bias $\Delta \mu$ were adjusted to reach the regime of maximum output power. As $\theta$ increases, the efficiency is lowered due to the increasing contributions of the undesired processes described in Fig. \ref{dragtrans}.
In the tight coupling regime, the thermodynamic efficiency attains values close to $\eta_{CA}$.

\begin{figure*}[htbp]
\centerline{\includegraphics[width=15.5cm]{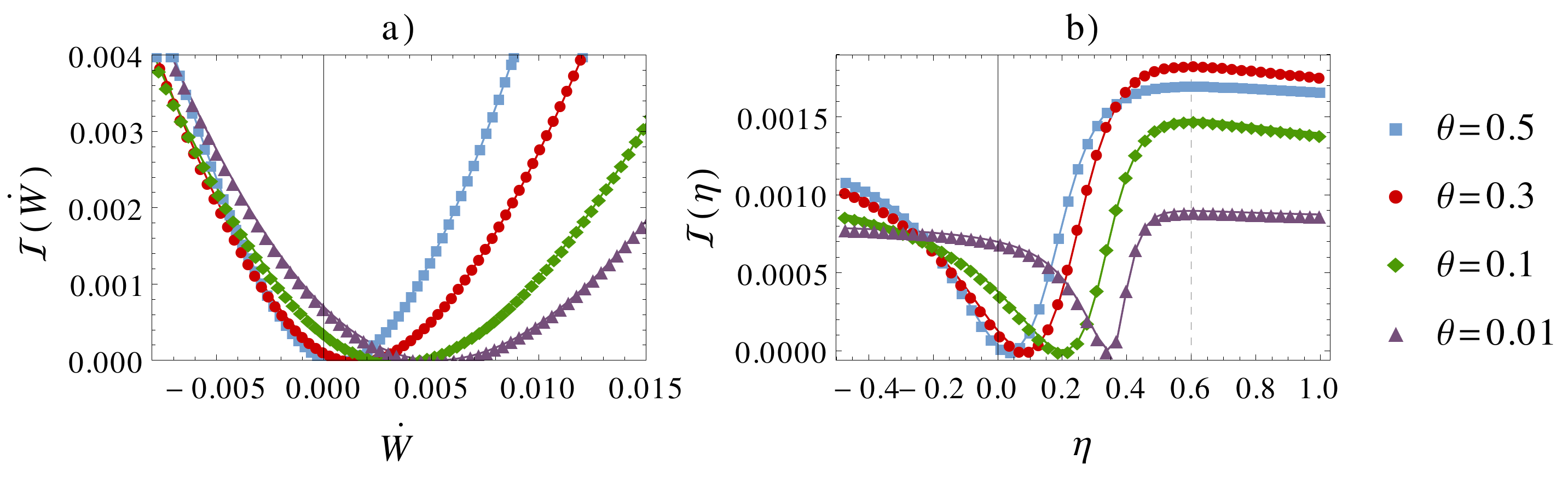}}
\caption{a) Output power LDF. b) Stochastic efficiency LDF. Parameters are chosen as in Fig. \ref{powerfig} b) and the Carnot efficiency is set to $\eta_C = 0.6$ in both Figs. These LDFs thus characterize the output power and efficiency fluctuations at the intersection points between the dashed vertical line and the curves illustrated in Fig. \ref{powerfig}. As explained in the text, the Carnot efficiency corresponds to the maximum of the LDF  (dashed vertical line) and is thus the least likely to be observed in a single run experiment.}
 \label{effmaxfig}
\end{figure*}

Up to this point we have considered the average properties of the output power and the so-called macroscopic efficiency, i.e. the efficiency defined as the ratio of the average output power over the average input power. In small devices displaying strong fluctuations, a more accurate characterization is provided by considering their statistical properties. Several works have recently shown that the so-called stochastic efficiency, defined along a single stochastic realisation of a thermal engine as
\be
\eta_s =  \frac{ \dot{\mathcal{W}}}{ \dot{\mathcal{Q}} },
\ee
exhibits universal properties \cite{Verley_2014_NatureCommunications, Rana_2014_ArXive-prints, Gingrich_2014_ArXive-prints, Verley_2014_ArXive-prints, Esposito_2015_arXivpreprintarXiv:1501.03232, Polettini_2015_Physicalreviewletters}. We now briefly show that the FCS of the currents developed in Sec. \ref{countingstatistics} can be used in order to study these fluctuations.

The probability distribution of a stochastic variable $\dot x = \Delta x/t $ is characterized in the long time limit by its large deviation function (LDF) $ \mathcal{I} (x)$, that is, $p(\Delta x) \sim \exp{\{-\mathcal{I} (\dot x) t\}}$ for $t \rightarrow \infty$. It is shown in Ref. \cite{Verley_2014_ArXive-prints} that the LDFs for the output power and stochastic efficiency LDFs can respectively be obtained as
\bea
\mathcal{I} (\dot \mathcal{W}) & = & \max_{\alpha} \left(  \mathcal{G}_{\dot w , \dot q} (\alpha , 0) - \alpha\dot \mathcal{W}  \right) \\
\mathcal{I } (\eta_s) & = & -\min_{\alpha} \mathcal{G}_{\dot w , \dot q} (\alpha \eta_s,  \alpha),
\eea
where the heat and work generating function $ \mathcal{G}_{\dot w , \dot q} (\alpha_w,  \alpha_q)$ is obtained by setting $\xi_3 = \xi_4 = \alpha_q$, $\lambda_1 = - \Delta \mu \alpha_w$ and the other counting parameters to $0$ in the current CGF (\ref{gendiffs}).

The LDFs of the fluctuating output power and efficiency are both illustrated on Fig. \ref{effmaxfig} a) and b) respectively, for different values of the asymmetry parameter chosen as in Fig. \ref{powerfig} and with the Carnot efficiency set to $\eta_C = 0.6$. Both output power and efficiency LDFs were evaluated in the regimes of maximal average output power. They thus characterize, respectively, the work fluctuations around the maxima's of the curves depicted on Fig. \ref{powerfig} a), and the efficiency fluctuations at the intersection points between the curves and the dashed vertical line in Fig. \ref{powerfig} b). 
The minimum of the output power LDF, corresponding to the most probable output power in the long time limit, corresponds to the average value of the output power which increases as one gets closer to tight coupling, i.e. as $\theta$ decreases. Similarly, the most likely value of efficiency in the long time limit, which lies the minimum of $\mathcal{I} (\eta_s)$, corresponds the macroscopic efficiency (\ref{effint6}) evaluated in Fig. \ref{powerfig} b). Furthermore, one observes that its maximum (along the dashed vertical line) lies at the Carnot efficiency ($\eta_C = 0.6$) in agreement with previous results \cite{Verley_2014_NatureCommunications, Rana_2014_ArXive-prints, Gingrich_2014_ArXive-prints, Verley_2014_ArXive-prints, Esposito_2015_arXivpreprintarXiv:1501.03232}. The Carnot efficiency is thus the least likely to be observed in the long time limit.


\subsection{Regime of isothermal electric current conversion}\label{motor}


We now consider an isothermal regime $\beta \equiv \beta_j$ $\forall j$, in which the current in the electrically biased QPC is converted into work done against the electrical bias applied to the DQD. The DQD and QPC channels are subject to the chemical potential biases
\be \label{entropyprosworkconv}
 \Delta \mu \equiv \mu_1 - \mu_2 <0 \\ \mbox{and}  \\ 0<  \Delta \mu_q \equiv \mu_3 - \mu_4 ,
\ee
respectively. The irreversible entropy production in the system at steady state is then given by (see Eq. (\ref{entropprod}))
\be
\dot{S}_i = \beta \left( -\langle \dot{\mathcal{W}}_{out} \rangle +\langle \dot{\mathcal{W}}_{in} \rangle \right) \geq 0
\ee
in terms of the input $\langle \dot{\mathcal{W}}_{in} \rangle = \Delta \mu_q \langle J_{N}^{3} \rangle$ and output $\langle \dot{\mathcal{W}}_{out} \rangle = - \Delta \mu \langle J_{N}^{1} \rangle$ power.

The efficiency of the conversion process can be written in terms of the input and output powers as
\be
0 \leq \eta \equiv \frac{ \langle \dot{\mathcal{W}}_{out} \rangle}{\langle \dot{ \mathcal{W}}_{in} \rangle} \leq 1.
\ee

\begin{figure*}[htbp]
\centerline{\includegraphics[width=11cm]{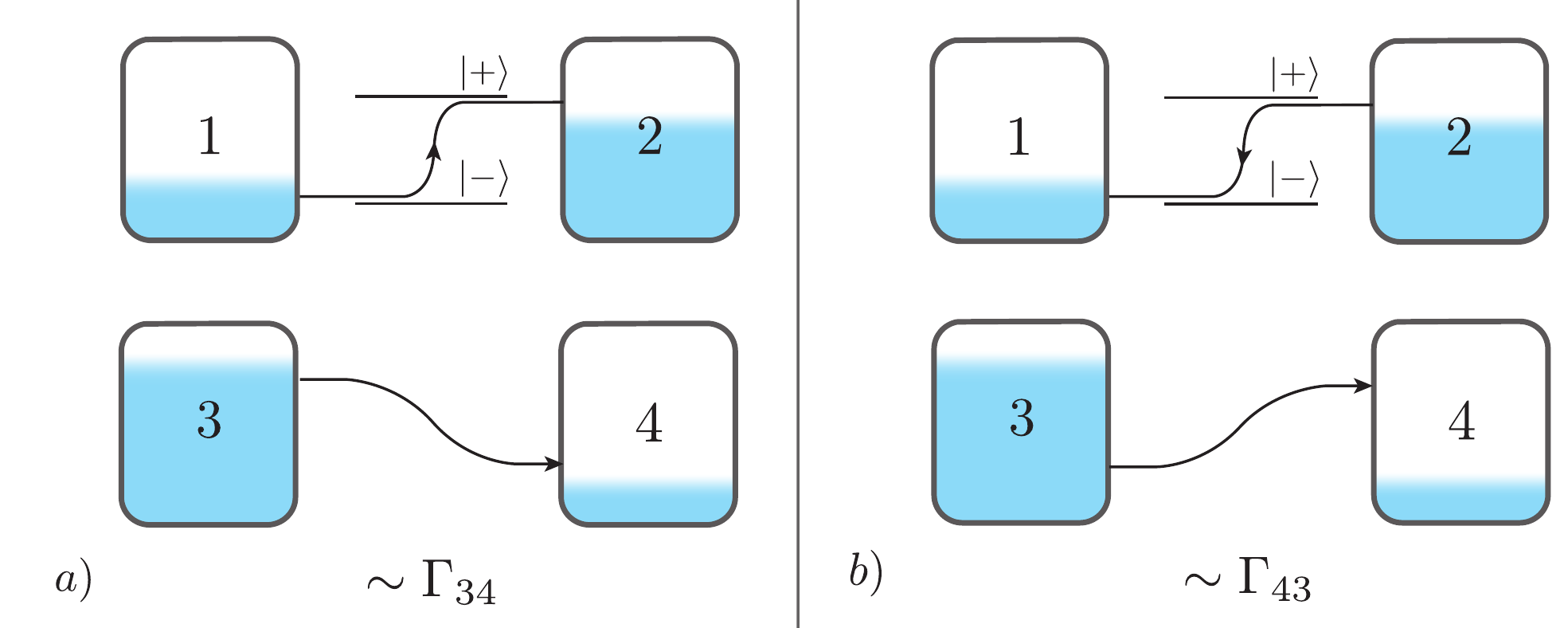}}
\caption{Illustration of the processes involving a transfer of a single electron from reservoir $3$ to $4$ and the corresponding transition in the DQD system.}
 \label{transitionsdrag3}
\end{figure*}

A strong asymmetry in the tunneling amplitudes of the DQD channel was needed in order to achieve high power and efficiency in the regime of thermoelectric conversion. In the present case, a similar asymmetry in the tunneling amplitudes of the QPC is also mandatory as a consequence of the directional nature of the driving processes in the QPC. To understand this point, it is useful to consider the low temperature limit of the machine. In this limit, energy is transferred from the QPC to the DQD only if $\Delta \mu_q > \omega_{+-}$. The several transfer processes involving a single electron transfer from reservoir $3$ to $4$ of the QPC, with associated transition in the DQD are depicted in Fig. \ref{transitionsdrag3}. By considering the rate constants involved in these two processes, we see that positive energy flow from the QPC to the DQD is enhanced provided
\be  \label{limitconv}
\Gamma_{43} \ll \Gamma_{34}
\ee
so that the process depicted in Fig. \ref{transitionsdrag3} a) becomes the dominant one.

On top of this, processes without net transfer of charge across the DQD like those depicted in Figs. \ref{dragtrans} b) and c) are also undesirable since they waist energy. We thus assume both (\ref{limitconv}) and (\ref{condmach}) to hold in order for our machine to work in optimal conditions. 

We further note that the DQD current and the contribution to the QPC current from electrons interacting with the DQD are tightly coupled in the regime where (\ref{condmach}) and (\ref{limitconv}) are satisfied, that is $\langle J_{N}^{1} \rangle \propto \langle J_{N}^{3} \rangle_d$. However, due to the presence of bare tunneling events in the QPC, the DQD current and the total QPC current are not necessarily tightly coupled. Only in the case $\gamma (\epsilon) = 0$ does the bare current in the QPC vanish, $\langle J_{N}^{3} \rangle_d = 0$, and the DQD current and the total current through the QPC become tightly coupled, i.e. $\langle J_{N}^{1} \rangle \propto \langle J_{N}^{3} \rangle_d = \langle J_{N}^{3} \rangle$.  

\begin{figure*}[htbp]
\centerline{\includegraphics[width=8cm]{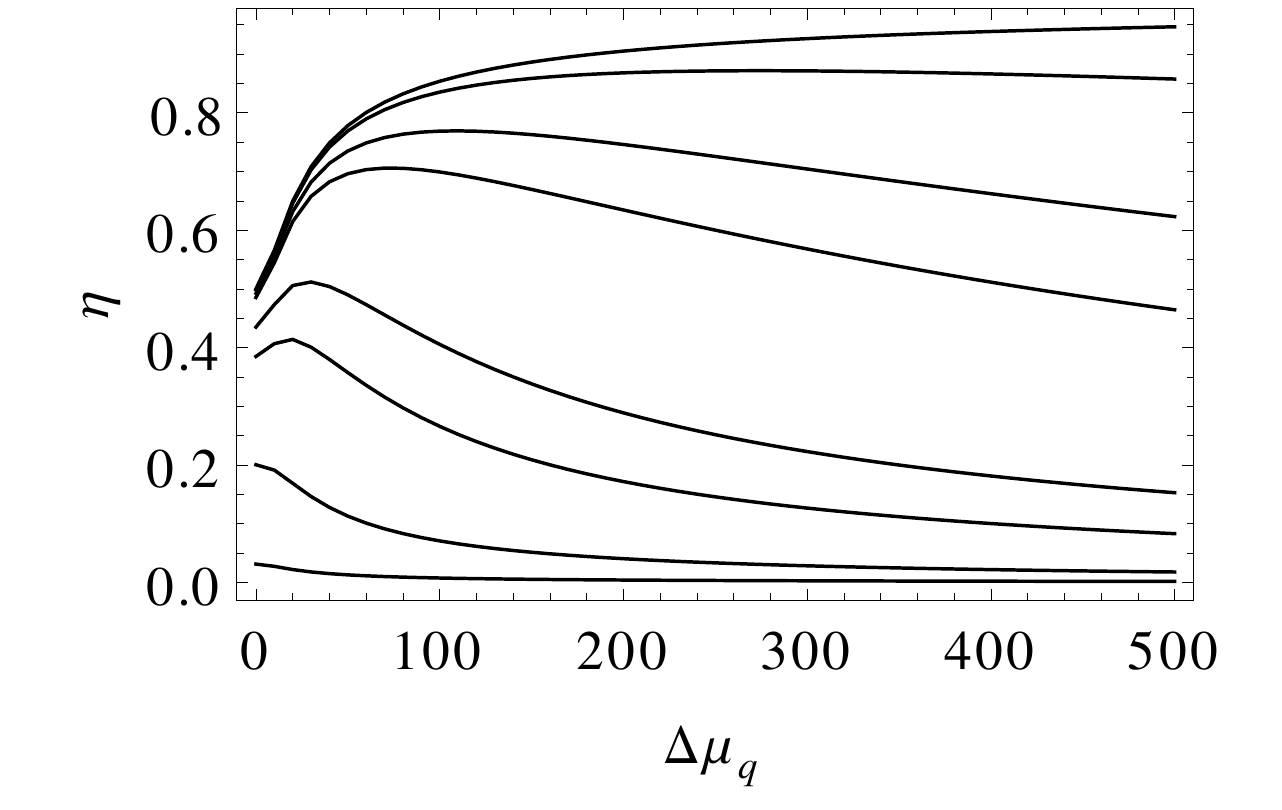}}
\caption{Efficiency of the current converter as a function of the input bias and for increasing bare tunneling parameter $\gamma$. From top to bottom line we chose $\gamma = 10^{-1}$, $5 \times 10^{-2}$, $10^{-2}$, $5 \times 10^{-3}$, $10^{-3}$, $5 \times 10^{-4}$, $10^{-4}$, $5 \times 10^{-5}$ and $10^{-5}$. The energies $E_+$ and $E_-$ and the bias $\Delta \mu$ in the DQD were optimized to reach maximum output power. The remaining parameters were chosen as $ \beta = 1$, $(V1+V2)/2 = 1.5$, $E_0 = 0$, $\Gamma_{2+} = \Gamma_{1-}= \Gamma_{34} =0.1$.}
 \label{effbias}
\end{figure*}

This remark has important consequences on the properties of efficiency at maximum power, and illustrates well the crucial role played by the FCS formalism in determining all the thermodynamically relevant processes when analysing thermodynamic machines. The efficiency is illustrated in Fig. \ref{effbias} for different values of the bare tunneling amplitude taken in the wide band limit, $\gamma (\epsilon) \equiv \gamma$. We observe a significant difference in the qualitative behavior of efficiency depending on the value of the bare tunneling amplitude $\gamma$. In particular, we observe that $\eta \rightarrow 1$ or $\eta \rightarrow 0$ in the far from equilibrium regime, i.e. $\Delta \mu_q \rightarrow \infty$, depending on whether $\gamma = 0$ or $\gamma \neq 0$ respectively. 

The behavior of efficiency is best understood by writing it as
\be \label{effworktowork}
\eta = \frac{ - \Delta \mu}{ \Delta \mu_q \left(1 + \frac{ \langle J_{N}^{3} \rangle_i }{ \langle J_{N}^{1} \rangle } \right) },
\ee
in terms of the bare current of electrons through the QPC $\langle J_{N}^{3} \rangle_i $ and where we assume both (\ref{condmach}) and (\ref{limitconv}).

For $\gamma = 0$, the bare current vanishes and the efficiency at maximum power in the equilibrium limit $\Delta \mu_q \rightarrow 0$ converges to the value $\eta = 1/2$, as predicted within linear response theory for systems working in the tight coupling regime \cite{VandenBroeck_2005_PhysicalReviewLetters}. In the large bias $\Delta \mu_q \rightarrow \infty$, efficiency reaches the maximal value $\eta = 1$. This is easily understood in the low temperature limit, i.e. $\beta \rightarrow \infty$. In this limit, the output power is non-zero provided that
\be \label{ineqcurconv}
\Delta \mu < \omega_{+-} < \Delta \mu_q.
\ee
When the bias $\Delta \mu_q$ applied to the QPC is large, the values of $\Delta \mu$ and $\omega_{+-}$ can be optimized to reach
\be
\frac{\Delta \mu}{\Delta \mu_q} \sim 1.
\ee

For $\gamma \neq 0$, a fraction of the electrons tunneling through the QPC dissipates entropy without exchanging energy with the DQD thus lowering the efficiency as suggested by (\ref{effworktowork}). Close to equilibrium, the efficiency converges to values below $\eta < 1/2$.

\begin{figure*}[htbp]
\centerline{\includegraphics[width=15cm]{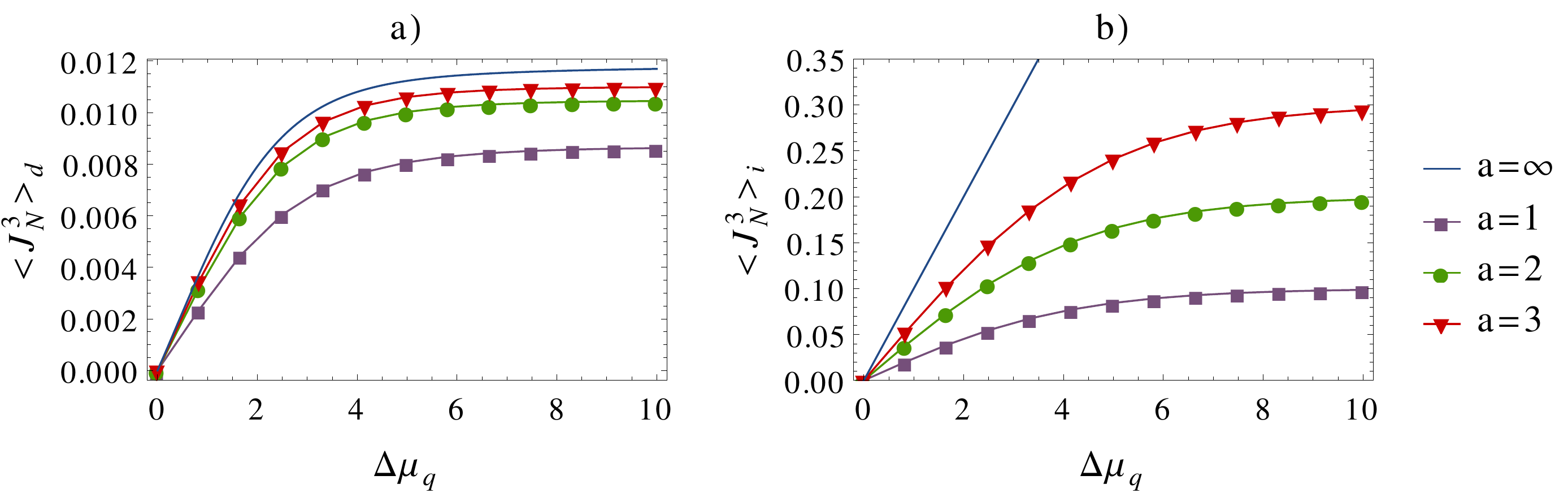}}
\caption{Curves of the current contributions in the QPC which do (left plot) and do not (right plot) induce transitions in the DQD, for different values of the band width $a$ (see text). Parameters have been chosen as $\beta=1$, $V_1 = V_2 = 1.5$, $E_0 = 0$, $E_+ = 0.5$, $E_- = 0.1$, $\Gamma_{2+} = \Gamma_{1-}= \Gamma_{34} = \gamma= 0.1$, }
 \label{currentsmach}
\end{figure*}

To understand the properties of efficiency in the large bias limit $\Delta \mu_q \rightarrow \infty$ we consider the behavior of the two contributions to the QPC current  $\langle J_N^{3}\rangle_d$ and $\langle J_N^{3}\rangle_i$ in Fig. \ref{currentsmach}. We have assumed the tunneling amplitude $\gamma (\epsilon)$, to be a stepwise function equal to $\gamma$ on the interval $\left[ -a/2 , a/2 \right]$ and $0$ elsewhere. In the wide band limit, i.e. $a \rightarrow \infty$, we see that the current $\langle J_{N}^{3} \rangle_i $ is linearly growing and diverges as $\Delta \mu_q \rightarrow \infty$ as can be seen from its definition (\ref{totcurr3})-(\ref{indcurrent}). This is in contrast to the current $\langle J_{N}^{3} \rangle_d$, which remains bounded $\forall \Delta \mu_q$ since it is ultimately constrained by the DQD splitting $\omega_{+-}$ as can be inferred from (\ref{totcurr3}), (\ref{curentdep}) and the expressions for the rates  (\ref{QPCrates6}) and (\ref{intratesQPC}).

These remarks together with expression (\ref{effworktowork}) for the efficiency and the fact that $\langle J_N^{1} \rangle$ remains bounded $\forall \Delta \mu_q$ explains why the efficiency decreases at least as fast as $\sim \Delta \mu_{q}^{-1}$ in the large bias limit $\Delta \mu_q \rightarrow \infty$ as soon as $\gamma \neq 0$.

We have thus shown that the bare tunneling events in the QPC can highly reduce the machine efficiency. The FCS formalism is here crucial in order to keep track of thermodynamically relevant processes which are otherwise missing in a stochastic description of the DQD populations. Let us finally mention that such processes did not affect the performance of the thermal engine considered in the present section due to the fact that the matter current in the QPC does not contribute to the entropy production since its corresponding affinity was set to zero, i.e. $\mu_3 = \mu_4$.


\section{Conclusion}\label{summary}


We fully characterized the nonequilibrium thermodynamics of a circuit composed of a double quantum dot (DQD) and quantum point contact (QPC) channels within the framework of stochastic thermodynamics. 

By using the modified quantum master equation formalism, we identified and provided a detailed description of all the microscopic processes contributing to the entropy changes in the system and the environment. We showed that the transition rates of processes related by time reversal satisfy the local detailed balance condition. This condition holds for the rates of electron transfers between the DQD and its reservoirs as well as for the rates of energy exchange processes with the out-of-equilibrium QPC. We also established a steady-state fluctuation theorem (FT) for the heat and matter currents across both channels, which reduces to a bivariate FT for the matter currents across each channel in an isothermal circuit.

Based on this analysis, we considered two regimes where the circuit operates as a thermoelectric or electric converter. We identified the optimal working  condition in both cases and evaluated the statistics of output power and efficiency at steady state.

Our study illustrates very well how stochastic thermodynamics enables one to structure the analysis of the transport properties of non-trivial mesoscopic devices such as the DQD-QPC circuit and to discriminate the universal (i.e. thermodynamic) features from the system specific ones. This theory has become an essential tool for quantum transport in the weak coupling regime. 


\section*{Acknowledgment}

The authors thank Pierre Gaspard for support and encouragement during this research.
This research was funded by the "Fonds pour la Formation à la Recherche dans 
l'Industrie et l'Agriculture" (FRIA Belgium) and by the National Research Fund, 
Luxembourg (project FNR/A11/02 and AFR Postdoc Grant 7982468).


\section*{References}




\section*{Diagonalisation of the DQD Hamiltonian}\label{basischange}


Diagonalisation of the DQD Hamiltonian (\ref{localbasisH}) can be performed analytically. We first introduce the basis of single dot occupation states as
\bea
| 1_A 0_B \rangle & = & c_{A}^{\dagger} | 0_A 0_B \rangle \\
| 0_A 1_B \rangle & = & c_{B}^{\dagger} | 0_A 0_B \rangle
\eea
where $| 0_A 0_B \rangle$ denotes the empty state of the DQD. The eigenstates of the DQD Hamiltonian (\ref{localbasisH}) are then given by
\bea
|0 \rangle & = &  | 0_A 0_B \rangle \\
| + \rangle & = & \cos \frac{\theta}{2} |1_A 0_B \rangle + \sin \frac{\theta}{2} |0_A 1_B \rangle \\
| - \rangle & = &  \sin \frac{\theta}{2} |1_A 0_B \rangle - \cos \frac{\theta}{2} |0_A 1_B \rangle
\eea
in terms of the mixing angle defined by
\be
\tan \theta = \frac{2T}{\epsilon_A - \epsilon_B}.
\ee
The corresponding energies are given in terms of the localised basis parameters by
\be
E_0 = 0 \\ 
E_{\pm} = \frac{\epsilon_{A} + \epsilon_{B} }{2} \pm \sqrt{\left(\frac{ \epsilon_{A} - \epsilon_{B}  }{2} \right)^{2} + T^2} .
\ee
Other parameters of the Hamiltonian in the single and manybody basis are related by
\bea
T_{1+}^{k} & = & t_{1A}^{k}  \cos \frac{\theta}{2} \\
T_{1-}^{k} & = & t_{1A}^{k}  \sin \frac{\theta}{2} \\
T_{2+}^{k} & = & t_{2B}^{k}  \sin \frac{\theta}{2} \\
T_{2-}^{k} & = & - t_{2B}^{k}  \cos \frac{\theta}{2} ,
\eea
and
\bea
T^{kk'}_{\pm \pm} & = & \frac{1}{2} (t^{kk'}_{A} + t^{kk'}_{B}) \pm  \frac{1}{2} (t^{kk'}_{A} - t^{kk'}_{B}) \cos \theta \\
T^{kk'}_{+-} & = & T^{kk'}_{-+} = \frac{1}{2} (t^{kk'}_{A} - t^{kk'}_{B}) \sin \theta.
\eea

\end{document}